\documentclass[aps,prd,twocolumn,eqsecnum,showpacs,amsmath]{revtex4}

\usepackage[dvips]{color,graphicx}
\usepackage{amsfonts,amssymb,theorem}
\newcommand{\D}{{\rm d}}

{\theorembodyfont{\upshape}
}
{\theorembodyfont{\upshape}
}
{\theorembodyfont{\upshape}
}
{\theorembodyfont{\upshape}
}
{\theorembodyfont{\upshape}
}
{\theorembodyfont{\upshape}
}

\newcommand{\dalm}{\kern1pt\vbox{\hrule height 0.9pt\hbox{\vrule width
0.9pt\hskip 2.5pt\vbox{\vskip 5.5pt}\hskip 3pt\vrule width 0.3pt}\hrule height
0.3pt}\kern1pt}

\begin{document}

\title{
The Roberts-(A)dS spacetime
}

\author{Hideki Maeda}
\email{h-maeda-at-hgu.jp}


\address{ 
Department of Electronics and Information Engineering, Hokkai-Gakuen University, Sapporo 062-8605, Japan
}

\date{\today}

\begin{abstract} 
Global structure of the (anti-)de~Sitter ((A)dS) generalization of the Roberts solution in general relativity with a massless scalar field and its topological generalization is clarified.
In the case with a negative cosmological constant, the spacetime is asymptotically locally AdS and it contains a black-hole event horizon depending on the parameters.
The spacetime may be attached to the exact AdS spacetime in a regular manner on a null hypersurface and the resulting spacetime represents gravitational collapse from a regular initial datum. 
The higher-dimensional counterpart of this Roberts-(A)dS solution with flat base manifold is also given.
\end{abstract}

\maketitle

\section{Introduction}

In 1989, Roberts presented an interesting exact solution in general relativity with a massless scalar field~\cite{roberts1989}.
This is a one-parameter family of dynamical and inhomogeneous solutions with spherical symmetry admitting a homothetic Killing vector.
An error of the expression has been corrected by several authors~\cite{sussman1991,ont1994,brady1994,burko1997} and a missing class of solutions in the original parametrization was also given~\cite{brady1994,hayward2000,ch2001}.
It was pointed out that, in the region where the derivative of the scalar field is timelike, this Roberts solution is equivalent to the solution obtained by Gutman and Bespal'ko for a stiff fluid in 1967~\cite{gb1967,maeda2009}.

The Roberts solution has been intensively investigated in the context of gravitational collapse.
The most fascinating feature of this solution is that the spacetime can be attached to the Minkowski spacetime on a null hypersurface in a regular manner, namely, without a massive thin shell.
Then the resulting spacetime represents gravitational collapse from a regular initial datum.
For this reason, the Roberts solution has been studied as a toy model to understand critical phenomena in gravitational collapse~\cite{ont1994,wo1997} or wormhole formation~\cite{maeda2009}.

In this background, it is natural to seek the (anti-)de~Sitter ((A)dS) generalization of the Roberts solution, but it has been missing for a long time.
Such a solution may be useful to understand the nature or final fate of the nonlinear turbulent instability of the AdS spacetime which was numerically found~\cite{br2011}.
Another possible application is the AdS/CFT duality~\cite{ads/cft} in the dynamical context.
The field theory at the boundary for a dynamical asymptotically AdS black hole should be in a non-equilibrium state and such a dynamical black hole has been constructed perturbatively as a holographic dual to the Bjorken flow~\cite{kmno}.
Indeed, there is an exact solution in the presence of a cosmological constant~\cite{lake1983,hajj1985,hp1994,vw1985,cl1987,sl1988}.
It represents an asymptotically locally AdS dynamical black hole with a homogeneous scalar field~\cite{maeda2012}.
However, it does not admit a limit to the Roberts solution.

Quite recently, Roberts himself successfully obtained the (A)dS generalization of his solution~\cite{roberts2014}.
This spacetime is conformally related to the Roberts spacetime and, interestingly, the configuration of the scalar field is totally the same as that in the solution without a cosmological constant.
Although this solution must have a variety of potentially interesting applications, only a few properties have been studied in~\cite{roberts2014}.
Especially, in order to know how useful it is, global structure of the spacetime must be clarified.

In this paper, we will present all the possible global structures of the Roberts-(A)dS solution and its topological generalization.
The paper is organized as follows.
In section II, we summarize basic properties of the solution.
All the possible Penrose diagrams are presented in section III and we summarize our results in the final section.
In appendix~\ref{app-A}, we present derivation of the Roberts-(A)dS solution and its higher-dimensional counterpart with $k=0$.
Our basic notation follows~\cite{wald}.
Greek indices run over all spacetime indices.
The convention for the Riemann curvature tensor is $[\nabla _\rho ,\nabla_\sigma]V^\mu ={R}^\mu _{~\nu\rho\sigma}V^\nu$ and ${R}_{\mu \nu }={R}^\rho_{~\mu \rho \nu }$.
The signature of the Minkowski metric is $(-,+,+,+)$ and we adopt the units of $c=1$.

\section{Preliminaries}
\subsection{System}
In the present paper, we consider the Einstein-$\Lambda$ system with a massless scalar field $\phi$ in four dimensions.
The field equations are
\begin{align}
&R_{\mu\nu}-\frac12g_{\mu\nu}R+\Lambda g_{\mu\nu} \nonumber \\
&~~~~~~=\kappa^2\biggl((\nabla_\mu\phi)(\nabla_\nu\phi)-\frac{1}{2}g_{\mu\nu}(\nabla \phi)^2\biggl),\\
&\dalm\phi=0,
\end{align}
where $\kappa:=\sqrt{8\pi G}$ and $(\nabla \phi)^2:=(\nabla_\rho \phi)(\nabla^\rho\phi)$.

We consider a warped product spacetime $(M^4,g_{\mu\nu}) \approx M^2\times K^{2}$, where $(M^2,g_{AB})$ is a two-dimensional Lorentzian manifold and $(K^{2},\gamma_{ij})$ is a two-dimensional unit space of constant curvature.
Indices $A,B$ take $0$ and $1$, while $i,j$ take $2$ and $3$.
The most general metric on such a spacetime is given by 
\begin{eqnarray}
\D s^2&=&g_{\mu\nu}\D x^\mu \D x^\nu \nonumber \\
&=&g_{AB}(y)\D y^A\D y^B+r(y)^2\gamma_{ij}(z)\D z^i\D z^j,\label{sol2}
\end{eqnarray}
where the warp factor $r$ is a scalar on $M^2$ which is interpreted as the areal radius.

The generalized Misner-Sharp quasi-local mass is a scalar on $M^2$ defined by 
\begin{align}
\label{qlm}
m := \frac{V_{(k)}}{\kappa^2}r\biggl(-\frac13\Lambda r^{2}+k-(D r)^2\biggl),
\end{align}  
where $D_A $ is the covariant derivative on $M^2$ and $(Dr)^2:=g^{AB}(D_Ar)(D_Br)$.
$k$ takes the values $1$, $0$, and $-1$, corresponding to positive, zero, and negative curvature of $K^{2}$, respectively~\cite{ms1964,nakao1995,maeda2006}.
Namely, the Riemann tensor on $K^{2}$, is given by 
\begin{eqnarray}
{}^{(2)}{R}{}_{ijkl}=k(\gamma_{ik}\gamma_{jl}-\gamma_{il}\gamma_{jk}).
\end{eqnarray}
$V_{(k)}$ denotes the volume of $K^{2}$ if it is compact.
In the spherically symmetric case, we have $V_{(1)}=4\pi$.

The generalized Misner-Sharp mass $m$ is constant in vacuum and it is zero for the maximally symmetric spacetime~\cite{hayward1996,mn2008}.
In addition, $m$ converges to the Arnowitt-Deser-Misner mass~\cite{ADM} and Abbott-Deser mass~\cite{AD} at spacelike infinity in the asymptotically flat and AdS spacetimes, respectively~\cite{nakao1995,hayward1996,mn2008}.

\subsection{Generalized Roberts-(A)dS solution}
In the recent paper~\cite{roberts2014}, Roberts presented a spherically symmetric solution in this system which is a (A)dS generalization of the Roberts solution in the system without $\Lambda$.
The topological generalization of this Roberts-(A)dS solution is given by 
\begin{align}
&\D s^2=\biggl(1-\frac{\Lambda}{6}uv\biggl)^{-2}\biggl(-2\D u\D v+S(u,v)^2\gamma_{ij}\D z^i\D z^j\biggl),\label{roberts-AdS}\\
&S(u,v)^2=-kuv+C_1v^2+C_2u^2,
\end{align}
where $C_1,C_2$ are constants and we have adopted different parametrization from the Roberts' paper.
(Derivation of this solution and its higher-dimensional counterpart with $k=0$ is presented in appendix~\ref{app-A}.)

For $k^2-4C_1C_2>0$, the scalar field $\phi$ is real and given by 
\begin{widetext}
\begin{align}
\label{phi-sol1}
\pm(\phi-\phi_0)=& \left\{
\begin{array}{ll}
\displaystyle{\frac{1}{\sqrt{2\kappa^2}}\ln\biggl|\frac{\sqrt{k^2-4C_1C_2}u+(ku-2C_1v)}{\sqrt{k^2-4C_1C_2}u-(ku-2C_1v)}\biggl|} & \mbox{for $C_1\ne0$},\\
\displaystyle{\frac{1}{\sqrt{2\kappa^2}}\ln\biggl|C_2-k\frac{v}{u}\biggl|} & \mbox{for $C_1=0$},
\end{array} \right. 
\end{align}
\end{widetext}
where $\phi_0$ is a constant.
For $k^2-4C_1C_2<0$, $\phi$ is ghost and given by
\begin{align}
\pm(\phi-\phi_0)=i\sqrt{\frac{2}{\kappa^2}}\arctan\biggl(\frac{ku-2C_1v}{\sqrt{4C_1C_2-k^2}u}\biggl). \label{phi-sol2}
\end{align}
If $k^2-4C_1C_2=0$, the field equations give $\phi=$constant and 
\begin{align}
R^{\mu\nu}_{~~\rho\sigma}=\frac{\Lambda}{3}(\delta^\mu_\rho\delta^\nu_\sigma-\delta^\mu_\sigma\delta^\nu_\rho),
\end{align}
namely, the spacetime is maximally symmetric.
In the case of $k^2-4C_1C_2=0$ with $k=\pm 1$, we have $S(u,v)^2=C_1(v-ku/2C_1)^2$ and hence $C_1$ and $C_2$ must be positive for physical solutions.
In the case of $k^2-4C_1C_2=0$ with $k=0$, $C_1=0$ with $C_2>0$ or $C_2=0$ and $C_1>0$ must be satisfied.

The expressions (\ref{phi-sol1}) and (\ref{phi-sol2}) give
\begin{align}
(\nabla\phi)^2=\frac{(k^2-4C_1C_2)uv(6-\Lambda uv)^2}{36\kappa^2(-kuv+C_1v^2+C_2u^2)^2}. \label{dphi2}
\end{align}
For a real scalar field, the derivative of  the scalar field is timelike, spacelike, and null in the regions with $uv<0$, $uv>0$, and $uv=0$, respectively.
Since a massless scalar field with timelike derivative is equivalent to a stiff fluid~\cite{madsen1988}, the regions with $uv<0$ can be described by the corresponding solution for a stiff fluid with a cosmological constant. (See Appendix A in~\cite{maeda2009}.)

The generalized Misner-Sharp mass (\ref{qlm}) for the Roberts-(A)dS spacetime is given by 
\begin{align}
m = &-\frac{3V_{(k)}(k^2-4C_1C_2)uv}{\kappa^2(6-\Lambda uv)\sqrt{-kuv+C_1v^2+C_2u^2}}.
\end{align}  
The spacetime is asymptotically (A)dS for $uv\to 6/\Lambda$ in the sense of
\begin{align}
\lim_{uv\to 6/\Lambda} R^{\mu\nu}_{~~\rho\sigma}\to \frac{\Lambda}{3}(\delta^\mu_\rho\delta^\nu_\sigma-\delta^\mu_\sigma\delta^\nu_\rho).
\end{align}
However, the generalized Misner-Sharp mass $m$ blows up in this limit.
Since $m$ converges to a constant at spacelike infinity in the asymptotically AdS spacetime~\cite{mn2008}, divergence of $m$ implies that the spacetime is only asymptotically locally AdS, namely the Henneaux-Teitelboim fall-off conditions to the AdS infinity~\cite{HT1985} are not satisfied.

In the case of $k^2-4C_1C_2>0$, the scalar field is real and $S(u,v)^2=0$ has real roots.
Then, another possible parametrization of the solution is 
\begin{align}
S(u,v)^2=\frac14(au-v)(bu-v),
\end{align}
where constants $a$ and $b$ satisfy $a+b=4k$, and then we have
\begin{align}
&m=-\frac{3V_{(k)}(a-b)^2uv}{8\kappa^2(6-\Lambda uv)\sqrt{(au-v)(bu-v)}},\\
&\pm(\phi-\phi_0)=\frac{1}{\sqrt{2\kappa^2}}\ln\biggl|\frac{bu-v}{au-v}\biggl|.
\end{align}
This is a similar parametrization to the one in the original paper~\cite{roberts2014}.
However, in this parametrization, we miss the solution with $C_1=C_2=0$, for example.

\section{Global structure of the Roberts-(A)dS spacetime}
In this section, we present all the possible global structures of the Roberts-(A)dS spacetime realized depending on the parameters $C_1$ and $C_2$.
The spacetime regions with $S(u,v)^2\le 0$ are unphysical because they do not have the spacetime signature $(-,+,+,+)$.

In the case of $k^2-4C_1C_2<0$, the scalar field is ghost and $C_1>0$ and $C_2>0$ are required for physical solutions holding $S(u,v)^2 \ge 0$.
The spacetime represents an interesting dynamical wormhole in this case, however, we will focus on the real scalar field in the present paper.
In the toroidal case ($k=0$), reality of the scalar field requires $C_1C_2<0$ and hence $C_1>0$ with $C_2<0$ and $C_1<0$ with $C_2>0$ are the only possibilities.

\subsection{Notes}
\begin{figure}[htbp]
\begin{center}
\includegraphics[width=0.6\linewidth]{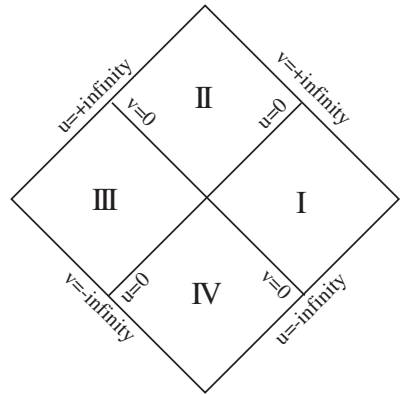}
\caption{\label{coordinate}
The Penrose diagram for the two-dimensional flat spacetime in the double null coordinates $u$ and $v$ ($-\infty<u,v<\infty$).
}
\end{center}
\end{figure}

In order to present the Penrose diagram, we take the transformations $u=\tan U$ and $v=\tan V$ in order to make coordinate infinities $u,v=\pm\infty$ being finite values.
Figure~\ref{coordinate} shows the Penrose diagram for the two-dimensional flat spacetime $\D s_2^2=-2\D u \D v$ in the double null coordinates $u$ and $v$ ranging from $-\infty$ to $\infty$.

In the flat case, the whole domain in Fig.~\ref{coordinate} represents one maximally extended spacetime.
In the case of the Roberts-(A)dS spacetime, in contrast, we will show that the whole domain in Fig.~\ref{coordinate} is divided into several portions by curvature singularities and null infinities.
Then, each potion corresponds to one distinct spacetime.

In addition, some of the portions do not represent maximally extended spacetimes because $u=\pm\infty$ and $v=\pm\infty$ are extendable boundaries, as shown below.

\subsection{Null infinity}
The Roberts-(A)dS spacetime admits the following conformal Killing vector:
\begin{align}
\xi^\mu\frac{\partial}{\partial x^\mu}=u\frac{\partial}{\partial u}+v\frac{\partial}{\partial v},
\end{align}
which satisfies 
\begin{align}
{\cal L}_\xi g_{\mu\nu}=&2\psi g_{\mu\nu},\\
\psi:=&\frac{6+\Lambda uv}{6-\Lambda uv}.
\end{align}
Therefore in this spacetime, there is a conserved quantity $C:=k_\mu \xi^\mu$ along null geodesics, where $k^\mu=\D x^\mu/\D \lambda$ is the tangent vector of a geodesic parametrized by an affine parameter $\lambda$. 

$\D u\D v=0$ is satisfied along radial null geodesics in this spacetime and so they are represented by $u=u_0$ or $v=v_0$, where $u_0$ and $v_0$ are constants.
Along $v=v_0$, the equation $C=k_\mu \xi^\mu$ is written as
\begin{align}
C=\biggl(1-\frac{\Lambda}{6}uv_0\biggl)^{-2}v_0\frac{\D u}{\D \lambda},
\end{align}
which is integrated to give
\begin{align}
\frac{\Lambda}{6}C(\lambda-\lambda_0)=\biggl(1-\frac{\Lambda}{6}v_0u\biggl)^{-1}, \label{null-u}
\end{align}
where $\lambda_0$ is an integration constant.
In a similar manner, we obtain
\begin{align}
\frac{\Lambda}{6}C(\lambda-\lambda_0)=\biggl(1-\frac{\Lambda}{6}u_0v\biggl)^{-1} \label{null-v} 
\end{align}
along $u=u_0$.
These equations show that $1-\Lambda v_0u/6=0$ or $1-\Lambda u_0v/6=0$ corresponds to $|\lambda|=\infty$, namely they are null infinity.

On the other hand, Eqs.~(\ref{null-u}) and (\ref{null-v}) show that $u\to \pm\infty$ along $v=v_0$ and $v\to \pm\infty$ along $u=u_0$ correspond to finite $\lambda$ and hence they are not infinity but extendable boundaries.
The extension of the spacetime beyond these boundaries will be studied later.

\subsection{Singularities}
Since the Ricci scalar of this spacetime is given by $R=4\Lambda+\kappa^2(\nabla\phi)^2$, Eq.~(\ref{dphi2}) shows that $S(u,v)^2=0$ gives curvature singularities unless $k^2-4C_1C_2=0$.
In addition, if $k\ne 0$, $v\to\pm\infty$ and $u\to\pm\infty$ are also curvature singularities for $C_1=0$ and $C_2=0$, respectively.

Let us study the curvature singularities given by $S(u,v)^2=-kuv+C_1v^2+C_2u^2=0$.
This equation describes two straight lines in the $(u,v)$-plane and their positions depend on $k$, $C_1$, and $C_2$.
For $k=0$, $C_1C_2<0$ is required for real scalar field and the singularities are given by
\begin{align}
u=\pm\sqrt{-\frac{C_1}{C_2}}v.
\end{align}
One is spacelike running in the regions I and III and the other is timelike running in the regions II and IV.

For $k=\pm 1$, the singularities are given by 
\begin{align}
v=0, \quad u=\frac{C_1}{k}v
\end{align}
for $C_2=0$ and 
\begin{align}
u=\frac{k\pm\sqrt{k^2-4C_1C_2}}{2C_2}v
\end{align}
for $C_2\ne 0$.
Hence, in the cases of $C_2=0$ with $kC_1<0$ and $C_1=0$ with $kC_2<0$ ($C_2=0$ with $kC_1>0$ and $C_1=0$ with $kC_2>0$), one singularity is null and the other runs in the regions I and III (II and IV).
If $C_2=C_1=0$, both singularities are null.
For $C_1C_2< 0$, one runs in the regions I and III and the other runs in the regions II and IV.
Both singularities run in the regions I and III (II and IV) in the cases of $k=1$ ($k=-1$) with $C_1<0$ and $C_2<0$ and $k=-1$ ($k=1$) with $C_1>0$ and $C_2>0$.

\subsection{Global structure of the spacetime}
We are now ready to present the Penrose diagrams for the Roberts-(A)dS spacetime.
Since the metric (\ref{roberts-AdS}) is invariant under the transformations $u\to -u$ and $v\to -v$, all the diagrams are centrally symmetric with respect to the origin $u=v=0$.

\begin{figure}[htbp]
\begin{center}
\includegraphics[width=1.0\linewidth]{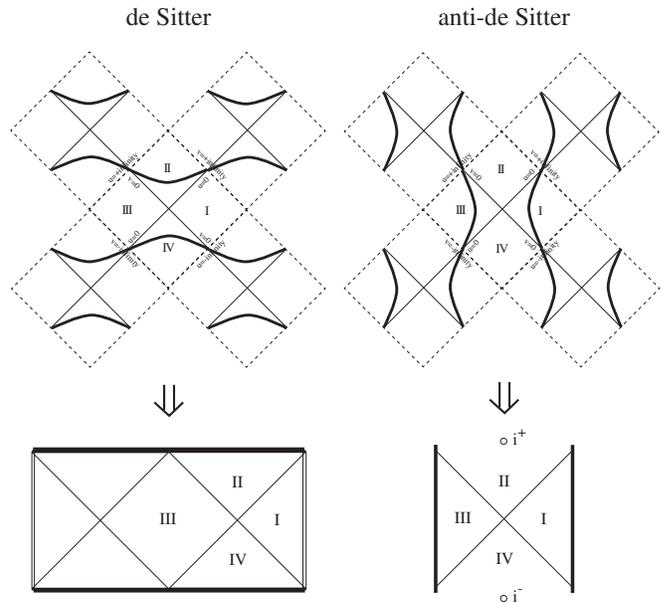}
\caption{\label{AdS}
The Penrose diagrams for the de~Sitter spacetime (left) and the anti-de~Sitter spacetime (right).
The same copies of the basic portion are attached at the dashed lines.
Thick curves correspond to the (A)dS infinity.
Each lower diagram is obtained from one maximally extended portion in the corresponding upper diagram.   
The double lines in the lower left diagram are identified~\cite{GP-text}.
$i^+$ and $i^-$ in the lower right diagrams are future and past timelike infinities, respectively.  
}
\end{center}
\end{figure}

As a lesson, we first present the Penrose diagram for the (A)dS spacetime (the case with $C_1C_2=k^2/4$) in Fig.~\ref{AdS}.
The well-known lower diagrams are given from a maximally extended portion in the upper diagrams.   

\begin{figure}[htbp]
\begin{center}
\includegraphics[width=0.9\linewidth]{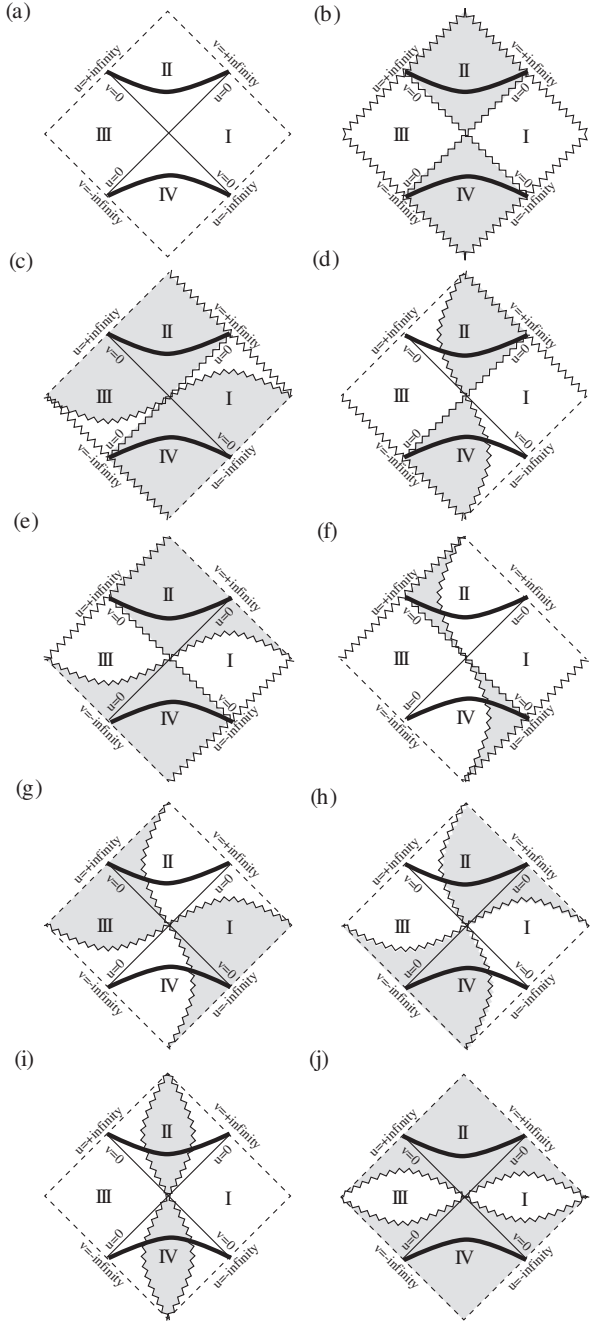}
\caption{\label{Penrose-dS}
All the possible Penrose diagrams for the Roberts-de~Sitter solution with a real scalar field.
A thick and a zigzag curve correspond to the dS infinity and a curvature singularity, respectively.
Dashed lines are extendable boundaries.
}
\end{center}
\end{figure}

\begin{figure}[htbp]
\begin{center}
\includegraphics[width=0.9\linewidth]{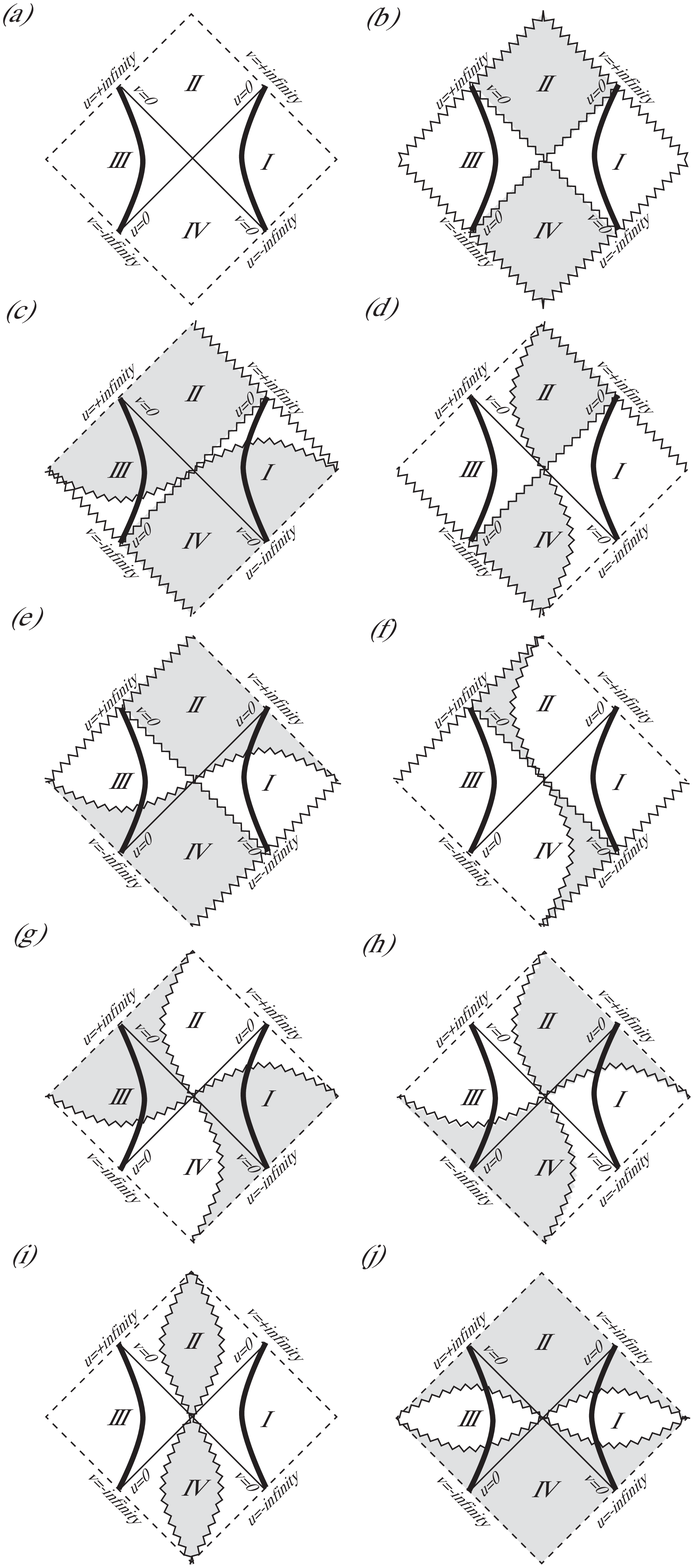}
\caption{\label{Penrose-AdS}
All the possible Penrose diagrams for the Roberts-anti-de~Sitter solution with a real scalar field.
A thick and a zigzag curve correspond to the AdS infinity and a curvature singularity, respectively.
Dashed lines are extendable boundaries.
}
\end{center}
\end{figure}

All the possible Penrose diagrams for the Roberts-(A)dS spacetime in the coordinates (\ref{roberts-AdS}) are presented in Figs.~\ref{Penrose-dS} and \ref{Penrose-AdS}.
The corresponding values of parameters $C_1$ and $C_2$ are summarized in Table~\ref{table:k}.
Here it is emphasized again that each position surrounded by curvature singularities and null infinities in one diagram corresponds one distinct spacetime.

In several cases, the coordinates (\ref{roberts-AdS}) do not cover the maximally extended spacetime.
If the extendable boundaries $u\to \pm\infty$ or $v\to \pm\infty$ are in the physical regions holding $S(u,v)^2>0$, we have to consider the spacetime extension beyond them in order to present the maximally extended spacetimes.

\begin{figure}[htbp]
\begin{center}
\includegraphics[width=1.0\linewidth]{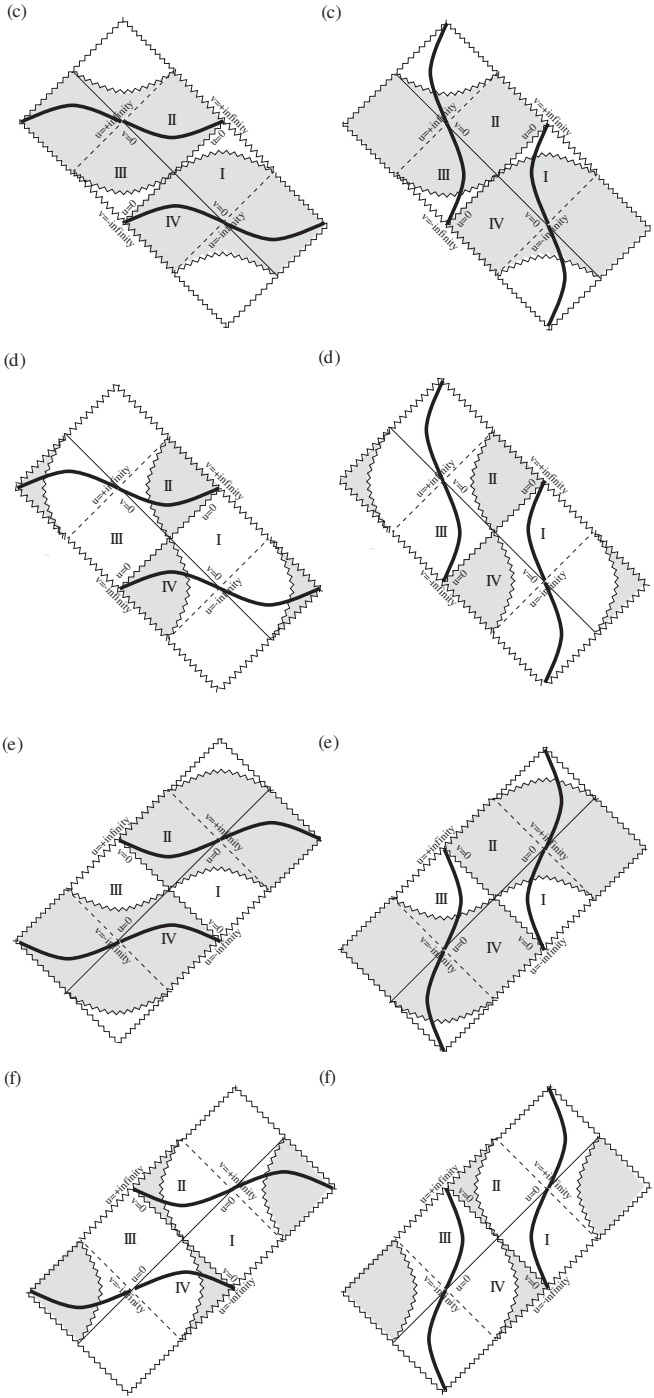}
\caption{\label{Roberts-AdS1}
Maximal extension of the spacetimes (c), (d), (e), and (f) in Figs.~\ref{Penrose-dS} (left) and \ref{Penrose-AdS} (right).
Each portion surrounded by thick and zigzag curves is one distinct maximally extended spacetime.
}
\end{center}
\end{figure}

\begin{figure}[htbp]
\begin{center}
\includegraphics[width=1.0\linewidth]{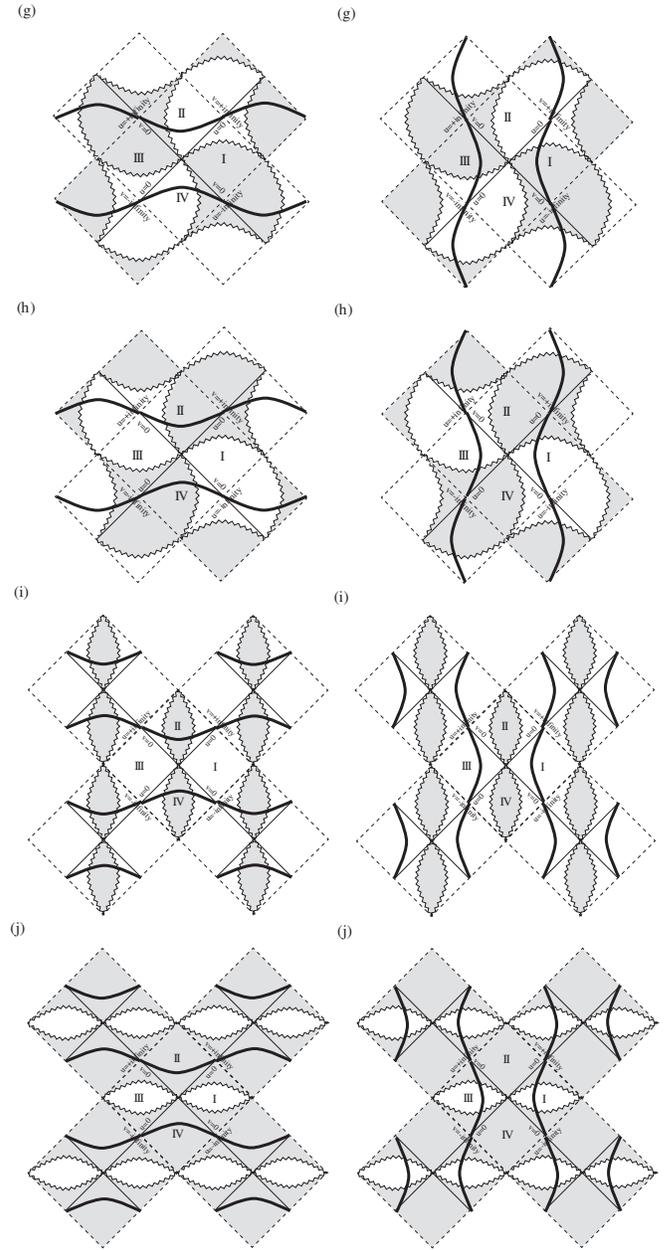}
\caption{\label{Roberts-AdS2}
Extension of the spacetimes (g), (h), (i), and (j) in Figs.~\ref{Penrose-dS} (left) and \ref{Penrose-AdS} (right).
Each portion surrounded by thick and zigzag curves is one distinct maximally extended spacetime.
Spacetime extension beyond the extendable boundaries (dashed lines) is performed in a similar manner.
}
\end{center}
\end{figure}

This extension is performed by the transformations $u=1/{\bar u}$ and $v=1/{\bar v}$.
The resulting metric is 
\begin{align}
&\D s^2=\biggl(\frac{\Lambda}{6}-{\bar u}{\bar v}\biggl)^{-2}\biggl(-2\D {\bar u}\D {\bar v}+{\bar S}({\bar u},{\bar v})^2\gamma_{ij}\D z^i\D z^j\biggl), \label{metric2} \\
&{\bar S}({\bar u},{\bar v})^2=-k{\bar u}{\bar v}+C_2{\bar v}^2+C_1{\bar u}^2,
\end{align}
which is non-singular at ${\bar u}=0$ or ${\bar v}=0$.
Causal structure of the spacetime covered by the above coordinates is the same as the original one with $u \leftrightarrow v$.
Finally, the maximally extended Roberts-(A)dS spacetimes are shown in Figs.~\ref{Roberts-AdS1} and \ref{Roberts-AdS2}.


\begin{center}
\begin{table}[h]
\caption{\label{table:k} Corresponding Penrose diagrams in Figs.~\ref{Penrose-dS} and \ref{Penrose-AdS} for the Roberts-(A)dS solution with a real scalar field depending on the parameters.
The unphysical (physical) regions holding $S^2<0$ ($S^2>0$) are in the shadowed regions for $k=1,0$ ($k=-1$). }

\begin{tabular}{l@{\qquad}c@{\qquad}c@{\qquad}c}
\hline \hline
  & $k=1$ & $k=0$ & $k=-1$   \\\hline
$C_1C_2=k^2/4$ & (a) & (a) & (a) \\ \hline
$C_1=C_2=0$ & (b) & n.a. & (b) \\ \hline
$C_1=0$, $C_2<0$ & (c) & n.a. & (d) \\ \hline
$C_1=0$, $C_2>0$ & (d) & (a) & (c) \\ \hline
$C_2=0$, $C_1<0$ & (e) & n.a. & (f) \\ \hline
$C_2=0$, $C_1>0$ & (f) & (a) & (e) \\ \hline
$C_1>0$, $C_2<0$ & (g) & (g) & (g) \\ \hline
$C_1<0$, $C_2>0$ & (h) & (h) & (h) \\ \hline
$0<C_1C_2<k^2/4$, $C_1>0$ & (i) & n.a. & (j) \\ \hline
$0<C_1C_2<k^2/4$, $C_1<0$ & (j) & n.a. & (i) \\ \hline
\hline
\end{tabular}
\end{table} 
\end{center}

\subsection{Attachment to the (A)dS spacetime}

Here we show that the Roberts-(A)dS spacetime can be attached without a massive thin shell to the (A)dS spacetime on a null hypersurface $u=0$ or $v=0$ and also $u=\pm \infty$ or $v=\pm\infty$ if they are regular.
(See~\cite{bi1991,Poisson} for the matching condition on a null hypersurface.)

Now $\Sigma$ denotes a matching null hypersurface $u=0$.
(The argument is the same for $v=0$.)
The induced metric $h_{ab}$ on $\Sigma$ is given by
\begin{eqnarray}
\D s_{\Sigma}^2=h_{ab}\D w^a \D w^b=C_1v^2\gamma_{ij}\D  z^i \D z^j,
\end{eqnarray}
where $w^a=(v,z^i)$ is a set of coordinates on $\Sigma$.
The basis vectors of $\Sigma$ defined by $e^\mu_a := \partial x^\mu/\partial w^a$ are 
\begin{align}
e^\mu_v\frac{\partial}{\partial x^\mu}&=\frac{\partial}{\partial v},\\
e^\mu_i\frac{\partial}{\partial x^\mu}&=\delta^\mu_{~i}\frac{\partial}{\partial z^i}
\end{align}
and the bases are completed by $N_\mu \D x^\mu=-\D v$.
They satisfy $N_\mu e^\mu_v=-1$ and $N_\mu e^\mu_i=0$ on $\Sigma$.
The only nonvanishing components of the transverse curvature $C_{ab}:=(\nabla_\nu N_{\mu}) e^\mu_{a} e^\nu_b$ of $\Sigma$ are
\begin{equation}
C_{ij}=\frac{v(C_1\Lambda v^2-3k)}{6} \gamma_{ij}.
\end{equation}

Regular attachment without a massive thin shell requires continuity of $h_{ab}$ and $C_{ab}$ at $\Sigma$.
Because $C_2$ does not appear in $h_{ab}$ and $C_{ab}$, two Roberts-(A)dS spacetimes with the same nonzero $C_1$ but different $C_2$ can be attached at $u=0$.

\begin{figure}[htbp]
\begin{center}
\includegraphics[width=0.4\linewidth]{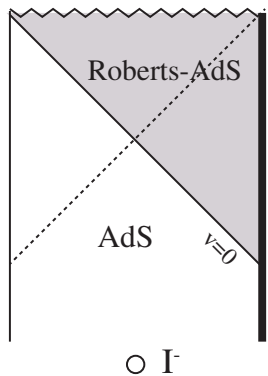}
\caption{\label{AdS-collapse-BH}
The Penrose diagram representing black-hole formation from a regular initial datum.
The Roberts-AdS spacetime (shadowed) given, for example, by the lower left portion in the region I in Fig.~\ref{Penrose-AdS}(h) is attached to the AdS spacetime on a null hypersurface $v=0$.
The dotted line represents a black-hole event horizon.
The left vertical line represents the symmetric center in the AdS spacetime.
}
\end{center}
\end{figure}

As a special case, the Roberts-(A)dS spacetime with $C_1={\bar C}_1(\ne 0)$ and $C_2={\bar C}_2$ can be attached to the (A)dS spacetime at $u=0$.
The parameters of this (A)dS spacetime are chosen such as $C_1={\bar C}_1$ and $C_2=k^2/(4{\bar C}_1) \ne {\bar C}_2$.
Similarly, two Roberts-(A)dS spacetimes with the same nonzero $C_2$ but different $C_1$ can be attached at $v=0$.
Attaching to the exact AdS spacetime in this manner, we can construct exact spacetimes representing black-hole or naked-singularity formation from a regular initial datum.
An example is shown in Fig.~\ref{AdS-collapse-BH}.

We can play the same game at $u=\pm\infty$ or $v=\pm \infty$ if they are regular.
For the proof, we use the metric (\ref{metric2}), where $u=\pm\infty$ and $v=\pm \infty$ correspond to ${\bar u}=0$ and ${\bar v}=0$, respectively.
We take a null hypersurface ${\bar u}=0$ as a matching surface $\Sigma$.
In this case, the induced metric on $\Sigma$ is 
\begin{eqnarray}
\D s_{\Sigma}^2=h_{ab}\D w^a \D w^b=\frac{36C_2}{\Lambda^2}{\bar v}^2\gamma_{ij}\D z^i\D z^j
\end{eqnarray}
and the nonvanishing components of the transverse curvature are
\begin{equation}
C_{ij}=\frac{{\bar v}(12C_2{\bar v}-k\Lambda)}{2\Lambda} \gamma_{ij}.
\end{equation}
Since $h_{ab}$ and $C_{ab}$ do not contain $C_1$, the spacetimes with different value of $C_1$ can be attached at $u=\pm\infty$.
In a similar manner, it is shown that the spacetimes with different value of $C_2$ can be attached at $v=\pm\infty$.

\section{Summary}
We have clarified all the possible global structures of the (A)dS generalization of the Roberts solution and its topological generalization.
The spacetime is conformally related to the Roberts spacetime and admits a conformal Killing vector.

While the Roberts spacetime in the double null coordinates represents a maximally extended spacetime, the coordinate infinity in the Roberts-(A)dS spacetime is a curvature singularity or a regular extendable boundary.
In the latter case, we have identified the extended regions of the spacetime and presented the Penrose diagrams for maximally extended spacetimes.
In the case with a negative cosmological constant, the spacetime is asymptotically locally AdS and it admits a black-hole event horizon depending on the parameters.

We have shown that the Roberts-(A)dS spacetimes with different parameters may be attached in a regular manner at coordinate origins or coordinate infinities if they are regular.
As a result, it is possible to construct exact spacetimes representing gravitational collapse from a regular initial datum.
They could be an interesting toy model of gravitational collapse of a massless scalar field in the presence of a cosmological constant.

In the context of the nonlinear instability of the AdS spacetime, dynamical stability of the Roberts-AdS solution is an important issue because the solution could describe the final state of the AdS instability if it is stable.
In the absence of $\Lambda$, the Roberts solution is stable against non-spherical linear perturbations~\cite{frolov1999} but has more than one unstable modes against spherical perturbations~\cite{frolov1997}.
Further studies of the Roberts-AdS solution are required to provide new insights on this problem, which will be reported elsewhere.

\subsection*{Acknowledgements}
The author thanks the anonymous referees for their careful reading of the manuscript and valuable comments, which significantly contributed to improving the quality of the publication.

\appendix

\section{Generalization of the Roberts-(A)dS solution}
\label{app-A}
In this appendix, we present the derivation of the Roberts-(A)dS solution and its higher-dimensional counterpart with $k=0$.
Let us consider the following $n$-dimensional metric and scalar field:
\begin{align}
\D s_n^2=&g_{\mu\nu}(x)\D x^\mu\D x^\nu \nonumber \\
=&-2\D u\D v+S(u,v)^2\gamma_{ij}\D z^i\D z^j,\\
\phi=&\phi(u,v),
\end{align}
where $\gamma_{ij}$ is the unit metric on the $(n-2)$-dimensional maximally symmetric space with its sectional curvature $k=1,0,-1$.
We assume that the functions $S(u,v)$ and $\phi(u,v)$ satisfy the Einstein equations $R_{\mu\nu}=\kappa_n^2(\nabla_\mu\phi)(\nabla_\nu\phi)$ and the Klein-Gordon equation $\dalm \phi=0$.

Now we consider the conformally related spacetime with the metric ${\bar g}_{\mu\nu}(x)=\Omega(x)^2g_{\mu\nu}(x)$ and assume that the new metric ${\bar g}_{\mu\nu}(x)$ and the same form of $\phi$ satisfy the field equations in the presence of a cosmological constant:
\begin{align}
{\bar R}_{\mu\nu}=&\kappa_n^2({\bar \nabla}_\mu\phi)({\bar \nabla}_\nu\phi)+\frac{2\Lambda}{n-2} {\bar g}_{\mu\nu},\label{beq1}\\
{\bar \dalm} \phi=&0.\label{beq2}
\end{align}
The Ricci tensor ${\bar R}_{\mu\nu}$ constructed from ${\bar g}_{\mu\nu}$ is written as
\begin{eqnarray}
{\bar R}_{\mu\nu}&=&R_{\mu\nu}-(n-2)\nabla_\mu \nabla_\nu\ln\Omega -g_{\mu\nu}\dalm \ln\Omega\nonumber \\
&&+(n-2)(\nabla_\mu\ln\Omega)(\nabla_\nu\ln\Omega) \nonumber  \\
&&-(n-2)g_{\mu\nu}(\nabla\ln\Omega)^2
\end{eqnarray}
and we have ${\bar \nabla}_\mu\phi=\nabla_\mu\phi$ and ${\bar \dalm} \phi=\Omega^{-2}\dalm\phi$.
Hence, Eqs.~(\ref{beq1}) and (\ref{beq2}) give the following equation for $\Omega$:
\begin{align}
&\nabla_\mu \nabla_\nu\ln\Omega+\frac{1}{n-2}g_{\mu\nu}\dalm \ln\Omega-(\nabla_\mu\ln\Omega)(\nabla_\nu\ln\Omega) \nonumber  \\
&+g_{\mu\nu}(\nabla\ln\Omega)^2=-\frac{2\Lambda}{(n-2)^2}\Omega^2 g_{\mu\nu}.\label{beq3}
\end{align}
Under an additional assumption $\Omega=\Omega(u,v)$, the above equation gives the following set of partial differential equations:
\begin{align}
&\partial_u\partial_u\ln\Omega=(\partial_u\ln\Omega)^2,\label{pde1}\\
&\partial_v\partial_v\ln\Omega=(\partial_v\ln\Omega)^2,\label{pde2}\\
&\partial_u\partial_v\ln\Omega=\frac{\Lambda}{(n-1)(n-2)}\Omega^2,\label{pde3}\\
&(\partial_u \ln S)(\partial_v\ln\Omega)+(\partial_v\ln S)(\partial_u\ln\Omega) \nonumber \\
&+(\partial_u\ln\Omega)(\partial_v\ln\Omega)=\partial_u\partial_v\ln\Omega.\label{pde4}
\end{align}

Equations (\ref{pde1})--(\ref{pde3}) can be solved without the information of $S(u,v)$ and the most general solution is
\begin{align}
\Omega(u,v)=&\left(p_0+p_1v+p_2u+p_3uv\right)^{-1},\label{Omega1}
\end{align}
where constants $p_0,\cdots,p_3$ satisfy
\begin{align}
p_1p_2-p_0p_3=\frac{\Lambda}{(n-1)(n-2)}.\label{p-lambda}
\end{align}

Lastly, we check whether Eqs.~(\ref{pde4}) is satisfied or not with the above $\Omega$ and the function $S(u,v)$ for the Roberts solution.
(See Appendix B in~\cite{maeda2012} for the expressions of $S(u,v)$ and also $\phi(u,v)$.)
In four dimensions, we have
\begin{align}
S(u,v)^2=-kuv+C_1v^2+C_2u^2,
\end{align}
with which Eqs.~(\ref{pde4}) gives the following constraints:
\begin{align}
2C_2p_1-kp_2=0,\quad 2C_1p_2-kp_1=0. \label{constraint}
\end{align}
In higher dimensions, the function $S(u,v)$ is obtained in a closed form only for $k=0$ as
\begin{align}
S(u,v)^2=(C_1v^{n-2}+C_2u^{n-2})^{2/(n-2)}.
\end{align}
With this expression, Eqs.~(\ref{pde4}) gives the same constraints (\ref{constraint}) with $k=0$.

We will see that $p_1p_2=0$ must be satisfied in all the cases above.
Then Eq.~(\ref{p-lambda}) reduces to
\begin{align}
p_0p_3=-\frac{\Lambda}{(n-1)(n-2)}
\end{align}
and $p_0$ can be set to be unity by rescaling transformations $u\to p_0u$ and $v\to p_0 v$. 

In the case of $k\ne 0$, constraints (\ref{constraint}) give
\begin{align}
p_2=\frac{2C_2p_1}{k},\quad (4C_1C_2-k^2)p_1=0.
\end{align}
Since $4C_1C_2-k^2 \ne 0$ is required for nontrivial solutions, we conclude $p_1=p_2=0$.
The Roberts-(A)dS solution corresponds to this case with $n=4$ and $k=1$.

In the case of $k=0$, there is more variety.
In this case, the constraints (\ref{constraint}) give
\begin{align}
C_2p_1=0,\quad C_1p_2=0.
\end{align}
Since $C_1=C_2=0$ is not allowed in this case, $p_1p_2=0$ is concluded.
If $C_1\ne 0$ and $C_2\ne 0$ hold, $p_1=p_2=0$ is satisfied.
If $C_1=0$ and $C_2\ne 0$ ($C_2=0$ and $C_1\ne 0$) hold, $p_1=0$ ($p_2=0$) is concluded.

\end{document}